\documentclass[runningheads,11pt]{llncs}

\usepackage[T1]{fontenc}
\usepackage{amsmath}
\usepackage[hidelinks]{hyperref}
\usepackage[sectionbib]{natbib}  
\usepackage{doi}
\usepackage{xcolor}
\usepackage{graphicx}
\usepackage{setspace} 
\usepackage{enumitem}
\usepackage[a4paper,
            bindingoffset=0.0cm,
            left=2.54cm,
            right=2.54cm,
            top=2.54cm,
            bottom=2.54cm,
            footskip=0.0cm]{geometry}
\usepackage{booktabs}
\usepackage{upgreek}
\newcommand{\dv}[1]{\boldsymbol{#1}}

\begin{document}

\title{On the sensitivity of different ensemble filters to the type of assimilated observation networks}
\titlerunning{On the sensitivity of different ensemble filters to the type of assimilated observation networks}

\author{Zixiang Xiong\inst{1}\thanks{This preprint has not been peer-reviewed and is shared by the authors to support the timely, noncommercial dissemination of research findings. Copyright remains with the authors, and the manuscript may not be copied or reposted without their explicit permission.} \and Siming Liang\inst{1}\href{https://orcid.org/0009-0002-9652-795X}{\includegraphics[scale=0.08]{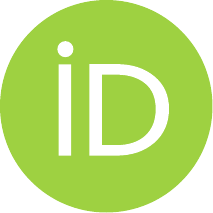}} \and Feng Bao\inst{1} \and \\ Guannan Zhang\inst{2}\href{https://orcid.org/0000-0001-7256-150X}{\includegraphics[scale=0.08]{orcid.pdf}} \and Hristo G.~Chipilski\inst{3}\href{https://orcid.org/0000-0003-3287-0038}{\includegraphics[scale=0.08]{orcid.pdf}}}
\authorrunning{Xiong, Liang, Bao, Zhang, and Chipilski}

\institute{Department of Mathematics, Florida State University, Tallahassee, FL \and Computer Science and Mathematics Division, Oak Ridge National Laboratory, Oak Ridge, TN \and Department of Scientific Computing, Florida State University, Tallahassee, FL \\ Corresponding author: \email{hchipilski@fsu.edu}}

\maketitle

\vspace*{-4mm}
\begin{abstract}
Recent advances in data assimilation (DA) have focused on developing more flexible approaches that can better accommodate nonlinearities in models and observations. However, it remains unclear how the performance of these advanced methods depends on the observation network characteristics. In this study, we present initial experiments with the surface quasi-geostrophic model, in which we compare a recently developed AI-based ensemble filter with the standard Local Ensemble Transform Kalman Filter (LETKF). Our results show that the analysis solutions respond differently to the number, spatial distribution, and nonlinear fraction of assimilated observations. We also find notable changes in the multiscale characteristics of the analysis errors. Given that standard DA techniques will be eventually replaced by more advanced methods, we hope this study sets the ground for future efforts to reassess the value of Earth observation systems in the context of newly emerging algorithms.

\end{abstract}


\section{Introduction}
\label{sec:intro}

Data assimilation (DA) methods have made steady progress in the geosciences ever since the early days of objective analysis schemes \citep{daley_1991,evensen_et_al_2021,kalnay_et_al_2024}. Advanced forms of ensemble and variational methods, along with hybrid approaches, now represent the state of the art in operational weather prediction centers \citep{bannister_2017}. These algorithms efficiently assimilate large volumes of observations to define the initial conditions of complex models, and enhance their forecast skill.

In parallel with these algorithmic advances, the Earth observing system has evolved significantly. For example, satellite observations are now routinely incorporated into models and have become a major contributor to improved analysis and forecast quality. In recent years, significant progress has been made in the assimilation of all-sky radiances \citep{geer_et_al_2018} and even data from alternative wavelengths, such as visible imagers \citep{scheck_et_al_2020}. At regional scales, assimilating high-frequency radar reflectivity and radial velocity data has demonstrated clear benefits for predicting convective storms \citep{dowell_et_al_2011,johnson_et_al_2015,gustafsson_et_al_2018,huang_et_al_2022}. Even in the historically under-sampled planetary boundary layer \citep{geerts2017}, novel profiling data from ground-based remote sensors \citep{wulfmeyer_et_al_2018,cimini_et_al_2020} and uncrewed autonomous systems \citep[UAS; ][]{barbieri_et_al_2019,bell_et_al_2020} present new opportunities to improve the short-term forecasts of high-impact weather events \citep{chipilski_et_al_2020,degelia2020,leuenberger_et_al_2020,jensen_et_al_2021,chipilski_et_al_2022,jensen_et_al_2022}.

The value of the global observation network is continuously monitored at operational centers running numerical weather prediction models. Prior to the deployment of new instruments, comprehensive testing is often performed to evaluate their added value, either through data denial experiments \citep{kelly_et_al_2007,baker_et_al_2022} or the generation of synthetic measurements using an Observing System Simulation Experiment (OSSE) framework \citep{arnold_dey_1986,masutani_et_al_2007,masutani_et_al_2010}. These efforts have greatly improved our understanding of how various observation system components enhance the numerical forecast skill. 

Nevertheless, one important limitation is that this understanding is largely tied to the standard ensemble and variational DA algorithms mentioned earlier. While these approaches are stable and computationally efficient for large geophysical applications, they are almost exclusively based on Gaussian assumptions. This limits their accuracy when applied to high-resolution models \citep{poterjoy_2022} or complex nonlinear observations, such as all-sky radiances \citep{geer_bauer_2011,minamide_zhang_2017}. By contrast, there has been growing interest in advanced non-Gaussian DA methods that offer more effective ways to blend sophisticated models and observations. These include non-Gaussian extensions of traditional methods \citep{amezcua_vanLeeuwen_2014,todter_et_al_2016,chan_et_al_2020,anderson_2022,nerger_2022,anderson_2023,fletcher_et_al_2023,chipilski_2025}, particle filter variants \citep{poterjoy_2016,poterjoy_et_al_2017,potthast_et_al_2019,vanLeeuwen_et_al_2019,rojahn_et_al_2023,hu_et_al_2024}, and data-driven (AI-based) approaches \citep{arcucci_et_al_2021,boudier_et_al_2023,bocquet_et_al_2024,wang_shen_2024,hammoud_et_al_2024,bach_et_al_2025,manshausen_et_al_2025}. The previously mentioned studies have already demonstrated the potential benefits of these advanced DA methods across a range of spatiotemporal scales. However, their sensitivity to the underlying observation network design has not been adequately explored.

The purpose of this study is to emphasize how the choice of an ensemble DA method (Gaussian vs.~non-Gaussian) can generate distinct analysis responses to the type of assimilated observations, especially in terms of their multiscale characteristics. These responses will be exemplified using an intermediate-complexity model and a non-Gaussian ensemble filter that has been recently developed by the authors \citep{EnSF_2023}. 

The presented numerical experiments follow some aspects of \citet{hamill_et_al_2002}, who also used a quasi-geostrophic model and computed the power spectrum of the background-error variances as a function of horizontal wavenumber. However, our work extends this framework in two important ways. First, we compare different ensemble DA methods, rather than evaluating a single approach (3D-Var in their case). Second, we enhance the scope of the conducted experiments: in addition to varying the number of assimilated observations, we also examine the effects of their (i) spatial distribution and (ii) type (linear vs.~nonlinear). Ultimately, we hope that insights from our experiments will inspire more studies that incorporate a broader set of traditional and non-Gaussian DA methods, as well as more realistic geophysical models.

\section{Methodology}
\label{sec:methodology}

To examine the sensitivity of different ensemble DA methods to the type of assimilated networks, we perform identical twin experiments which fit into the OSSE framework mentioned earlier. In this framework, synthetic observations are generated using the same numerical model that is later employed in the DA experiments \citep{kao_et_al_2004}.

\subsection{Numerical model}

For this study, we use the surface quasi-geostrophic (SQG) model, which exhibits turbulent characteristics similar to those of real geophysical flows \citep{rotunno_snyder_2008}. The SQG system assumes uniform potential vorticity, which is bounded between two horizontal surfaces separated by 10~km. We adopt the formulation of \citet{tulloch_smith_2009} which captures the nonlinear Eady dynamics. In this setup, the governing equations reduce to the advection of potential temperature on the two bounding surfaces. 

These equations are integrated in time using a fourth-order Runge-Kutta scheme with a 20-minute time step. Spatial derivatives are numerically approximated using a spectral collocation method based on the fast Fourier transform, taking advantage of the doubly periodic boundary conditions. All experiments are performed on a $64 \times 64$ grid on each of the two model surfaces, resulting in a model state dimension of 8192. Additional details on the model configuration can be found in \citet{wang_et_al_2021} and in the GitHub repository dedicated to this paper (\url{https://github.com/ZixiangXiong/sqgpublic}).

\subsection{Observation networks}

To perform identical twin DA experiments, we first initialize the SQG model with standard Gaussian noise and impose a sinusoidal pattern on the upper model surface. The SQG model is integrated for 300 days, with the first 100 days discarded to allow for sufficient model spin-up. The nature run (i.e., the true simulation) is defined as the model trajectory between days 100 and 150. During this period, synthetic observations are generated every 3 hours, resulting in a total of 400 observation times. The observation generation process at a given time can be described with stochastic model
\begin{equation}
\dv{\bf{y}} = \dv{\bf{h}}(\dv{\bf{x}}_t) + \dv{\upeta},\quad \text{with } \dv{\upeta} \sim \mathcal{N}(\dv{0},\dv{\bf{R}}),
\end{equation}
where $\dv{\bf{x}}_t$ is the true SQG state, $\dv{\bf{h}}$ is the observation operator, and $\dv{\upeta}$ is a Gaussian observation error realization with a zero mean and covariance $\dv{\bf{R}}$. We consider two observation types:

\setlength{\leftmargini}{3.5mm}
\begin{itemize}[label=\textbullet]
  \setlength\itemsep{2mm}
  \item \textit{Linear}: The observation operator $\dv{\bf{h}}$ is a selection matrix that identifies a set of observed grid points. The true SQG values are then corrupted by Gaussian noise with covariance $\dv{\bf{R}} = \dv{\bf{I}}$.
  \item \textit{Nonlinear}: The operator $\dv{\bf{h}}$ indexes a subset of the observed grid points and applies a nonlinear arctangent function. For this nonlinear observation type, we use $\dv{\bf{R}} = 0.01^2 \times \dv{\bf{I}}$.
\end{itemize}
\vspace*{-4mm}

\begin{figure}[ht!] 
\centering
\includegraphics[width=0.75\textwidth]{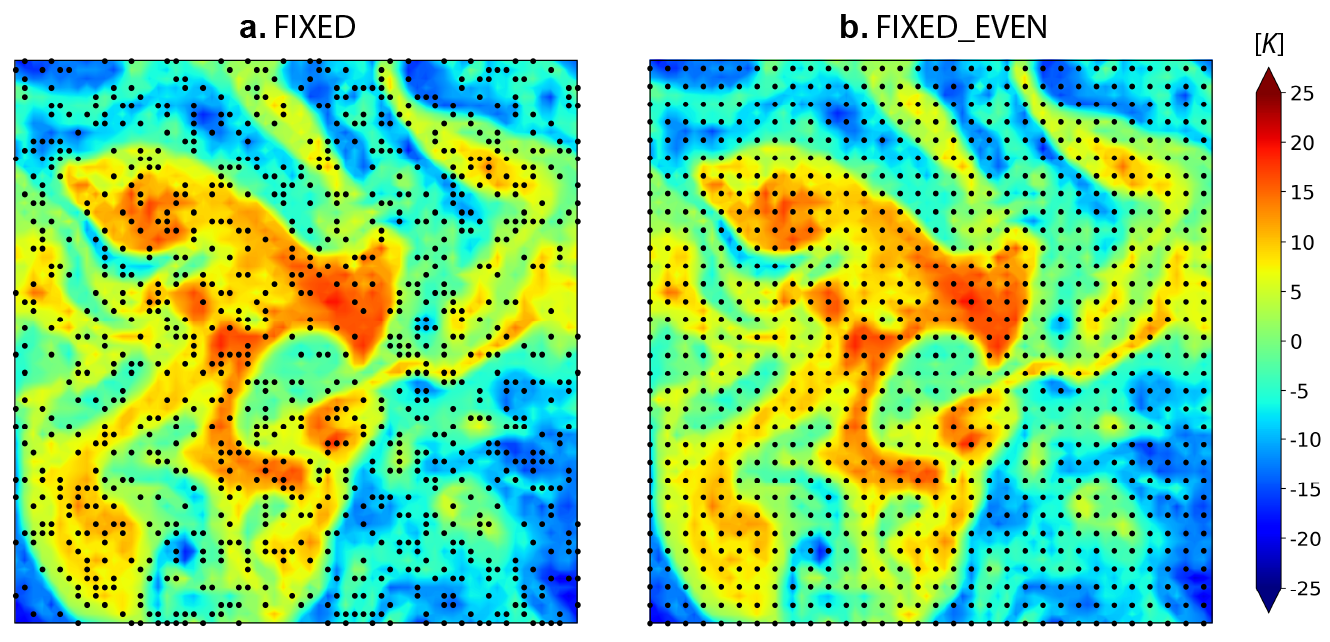}
\caption{Observation locations in the FIXED and FIXED\_EVEN networks (black dots), overlaid on an example realization of the potential temperature anomaly from the nature run (color shading).}
\label{fig1_obNetworks}
\end{figure}

In addition to varying the percentage of nonlinear observations (see Table~\ref{tab:letkf_vs_ensf}), we also consider 3 observation networks with different spatial distributions, closely following the study of \citet{morss_emanuel_2001}. In the first two, observation locations are fixed in time: FIXED\_EVEN uses evenly spaced observations, while FIXED randomly selects the observed grid points at the start of each DA experiment, keeping the same locations during the entire simulation. Figure~\ref{fig1_obNetworks} shows example distributions with 1024 observations from these two networks, overlaid on an example realization of the SQG state. The third network, referred to as RANDOM, has a spatial distribution similar to FIXED (Figure \ref{fig1_obNetworks}a) but changes the observation locations every analysis time. Thus, the RANDOM network can be viewed as a naive or baseline adaptive sampling strategy, which has not been specifically optimized to reduce the forecast error growth \citep{morss_emanuel_2001}. 

\subsection{Ensemble data assimilation methods}

The initial ensemble for our DA experiments is formed by randomly selecting 20 states from the last 200 days of the true SQG integration (which spans a total of 300 days; see earlier discussion). The ensemble filters are evaluated over the same 50-day period during which the synthetic observations were generated, resulting in 400 analysis cycles. The primary objective of the numerical experiments presented in the next section is to evaluate the impact of different observation networks on both standard and recently developed ensemble filters. 

For our reference DA approach, we adopt the Local Ensemble Transform Kalman Filter \citep[LETKF; ][]{hunt_et_al_2007}, which relies on Gaussian assumptions during the analysis step. Despite these limitations, LETKF's efficient local-domain implementation has made it a widely used EnKF variant, including its operational use in several NWP centers \citep{schraff_et_al_2016,frolov_et_al_2024}.

To assess how more advanced nonlinear ensemble filters respond to changes in the assimilated observation networks, we also consider the Ensemble Score Filter \citep[EnSF; ][]{EnSF_2023,EnSF_SQG}, which employs a training-free diffusion process to accommodate for non-Gaussian prior distributions and nonlinear observations. Although other promising nonlinear algorithms were reviewed earlier, we focus on EnSF because our recent work has demonstrated its ability to provide stable filtering results for the SQG system \citep{EnSF_SQG}. This makes it a compelling candidate for exploring the sensitivity of nonlinear filters to the various observation networks considered in this study.

\section{Results}
\label{sec:results}

Since the primary advantage of EnSF over LETKF lies in its ability to handle nonlinearities in the analysis step, we first examine the impact of increasing the fraction of nonlinear observations. Specifically, we consider a scenario in which 25\% of the SQG model state is observed (1024 observations in total) and progressively increase the proportion of arctangent observations in increments of 20\% across all three observation networks.

In the fully linear case (0\% arctangent observations), Table \ref{tab:letkf_vs_ensf} shows that LETKF is nearly four times more accurate across all networks. This finding aligns with our recent results in \citet{EnSF_SQG}, which demonstrated that a well-tuned LETKF remains advantageous in an idealized, fully linear setup. Across different networks, we observe that the largest analysis mean errors occur in the FIXED network, whereas FIXED\_EVEN and RANDOM achieve comparable performance. This pattern holds for both LETKF and EnSF, underscoring the importance of evenly sampling geophysical systems characterized by well-developed turbulence.

For all hybrid (linear-nonlinear) observation networks in Table \ref{tab:letkf_vs_ensf}, EnSF consistently yields visibly lower analysis RMSEs. Notably, the relative impacts of different networks remain unchanged: FIXED continues to yield the worst performance, while RANDOM and FIXED\_EVEN achieve similar results. However, in all nonlinear LETKF experiments, FIXED\_EVEN emerges as the optimal network, consistently outperforming RANDOM. The latter ranks as the second-best network in most nonlinear experiments, except in the case with 80\% arctangent observations, where FIXED proves more beneficial.

\begin{table}[ht!]
\centering
\caption{Time-averaged analysis RMSEs from LETKF and EnSF experiments assimilating 1024 observations (25\% of all state variables). Each row represents a different percentage of nonlinear (arctangent) observations in the system.}
\begin{tabular}{c @{\hspace{1.2cm}} c@{\hspace{4mm}} c@{\hspace{4mm}} c @{\hspace{1.2cm}}  c@{\hspace{4mm}} c@{\hspace{4mm}} c}
\toprule
 &\multicolumn{3}{c}{LETKF} & \multicolumn{3}{c}{EnSF}\\[1mm]
Arctan\% & \small{FIXED} & \small{FIXED\_EVEN} & \small{RANDOM} & \small{FIXED} & \small{FIXED\_EVEN} & \small{RANDOM}\\
\hline
0\% & 0.716 & 0.642 & 0.645 & 2.756 & 2.470 & 2.425 \\
20\% & 10.117 & 7.957 & 9.015 & 2.780 & 2.482 & 2.446 \\
40\% & 9.885 & 8.967 & 9.086 & 2.792 & 2.516 & 2.445 \\
60\% & 9.632 & 9.531 & 9.695 & 2.797 & 2.508 & 2.458 \\
80\% & 9.440 & 9.083 & 10.128 & 2.798 & 2.515 & 2.471 \\
100\% & 9.209 & 9.044 & 9.175 & 2.836 & 2.551 & 2.501 \\
\bottomrule
\end{tabular}
\label{tab:letkf_vs_ensf}
\end{table}

\begin{figure}[ht!] 
\centering
\includegraphics[width=0.85\textwidth]{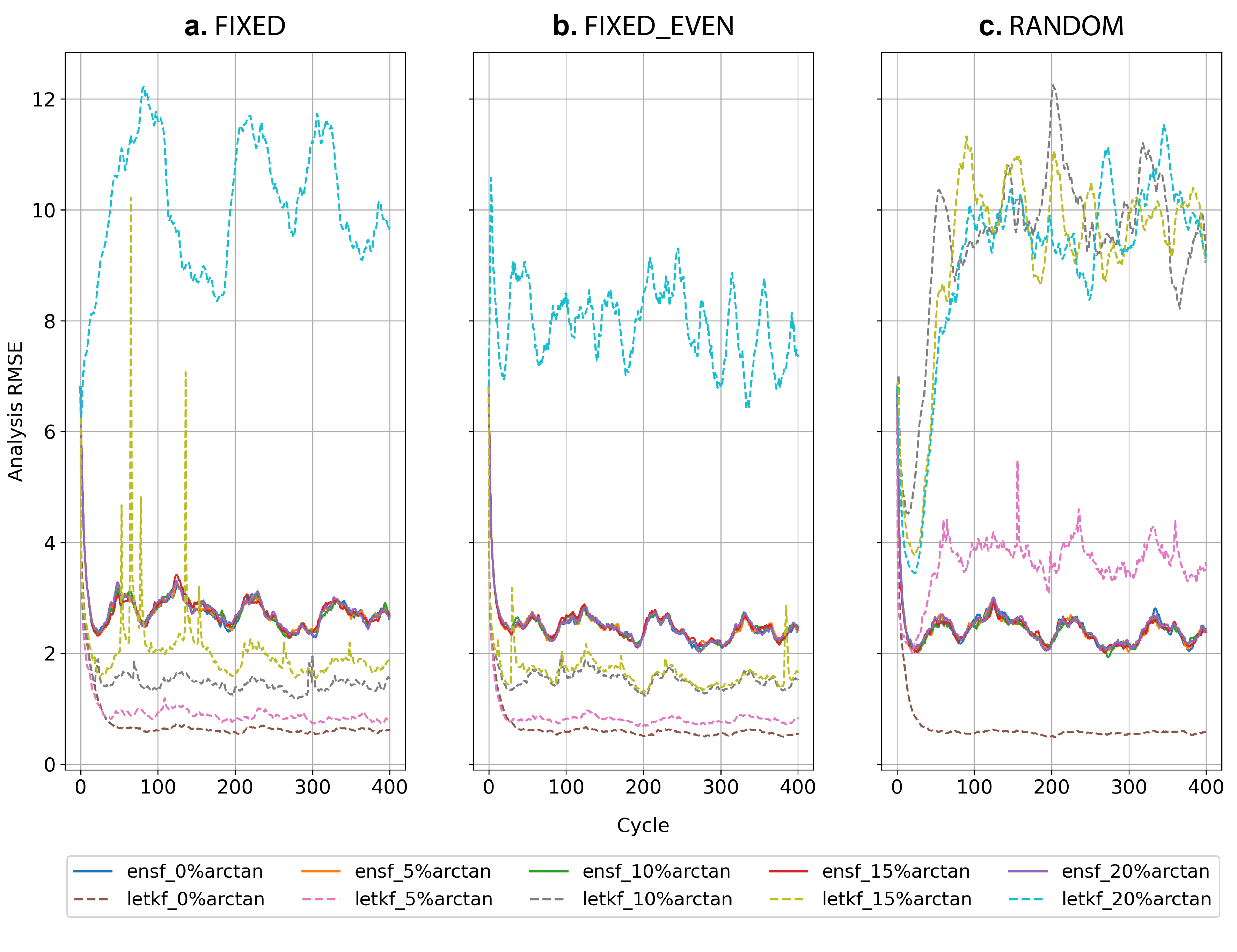}
\vspace*{-5mm}\caption{Evolution of the analysis RMSEs for LETKF (dashed lines) and EnSF (solid lines) with the FIXED, FIXED\_EVEN and RANDOM networks, each assimilating 1024 observations (25\% of all state variables). Different colors indicate varying percentages of nonlinear (arctangent) observations.}
\label{fig2_rmse_cycle}
\end{figure}

The previous results indicate a shift in the behavior of the two ensemble filters for nonlinearities between 0\% and 20\%. This raises the question of identifying the threshold fraction of nonlinear observations at which LETKF begins to degrade. To address this, Figure \ref{fig2_rmse_cycle} presents a detailed evolution of the analysis mean errors as the fraction of arctangent observations increases from 0\% to 20\% in increments of 5\%.

The first key observation from the analysis RMSE time series in Figure \ref{fig2_rmse_cycle} is that EnSF maintains consistent performance regardless of the degree of nonlinearity. On the other hand, the LETKF experiments confirm that the FIXED\_EVEN network remains optimal in the presence of nonlinearities. With 20\% arctangent observations, the RANDOM network slightly outperforms FIXED, as shown in Table \ref{tab:letkf_vs_ensf}. However, a new insight from the time series is that the RANDOM network results in a more rapid deterioration of LETKF's analysis accuracy. While LETKF retains its advantage over EnSF up to 15\% observational nonlinearity in the FIXED and FIXED\_EVEN networks, its performance in the RANDOM network begins to degrade as early as 5\% arctangent observations.

\begin{figure}[ht!] 
\centering
\includegraphics[width=0.8\textwidth]{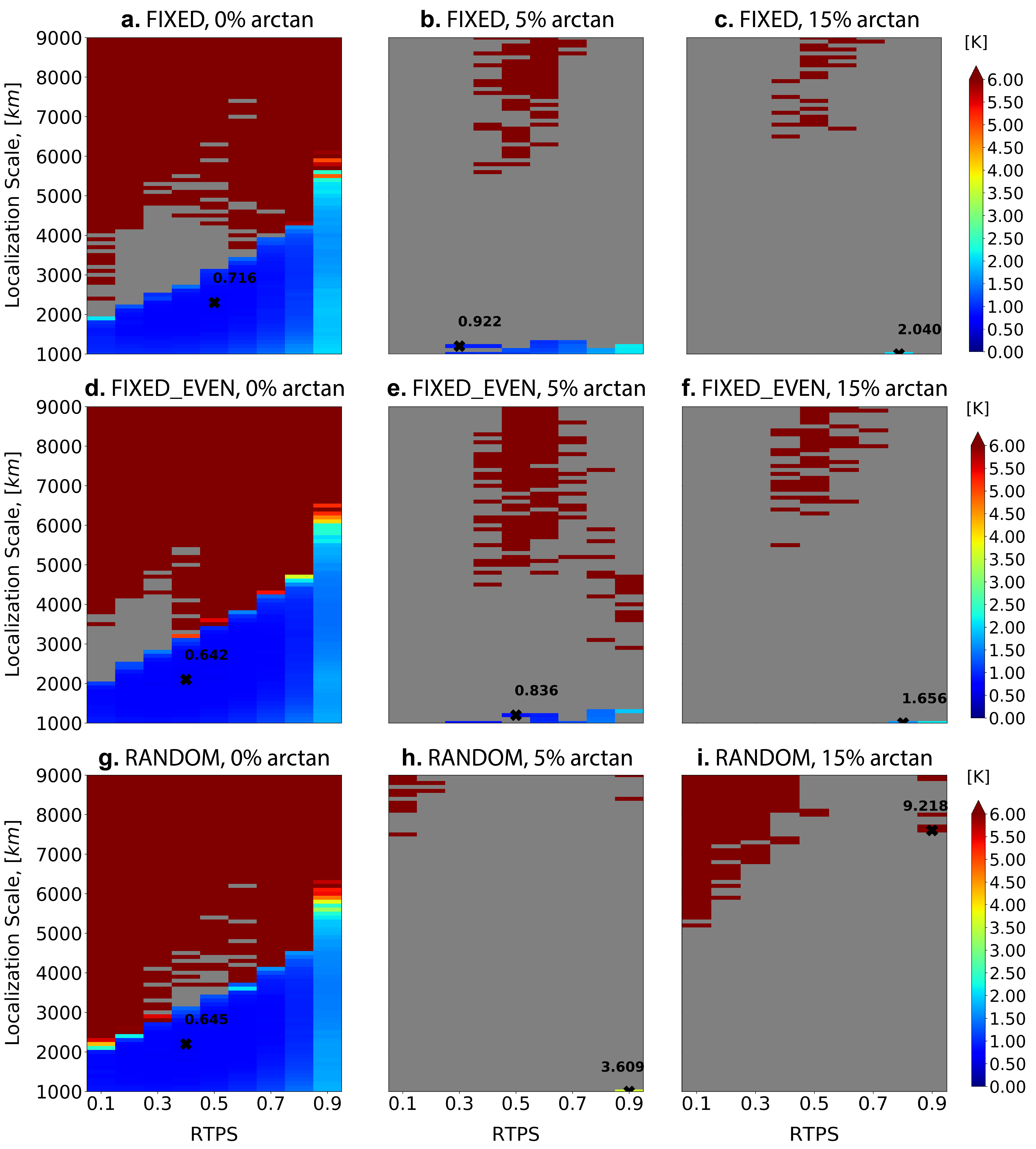}
\caption{Illustrating the challenges in optimally tuning the LETKF algorithm as the number of nonlinear (arctangent) observations increases. Each plot shows the time-averaged analysis RMSEs as a function of the RTPS (relaxation to prior spread) inflation parameter and horizontal localization scale. Rows correspond to different observation distributions in space (FIXED, FIXED\_EVEN and RANDOM), while columns indicate varying percentages of nonlinear (arctangent) observations. Similar to Figure \ref{fig2_rmse_cycle}, the total number of assimilated observations is 1024 (25\% of all state variables).}
\label{fig3_letkf_tuning}
\end{figure}

Another consequence of increasing the degree of nonlinearity in the observing system is the growing difficulty of determining optimal tuning parameters for the benchmark LETKF algorithm. Figure \ref{fig3_letkf_tuning} shows that in a fully linear observation network, there exist a broad range of localization and inflation parameters for which LETKF performs well (indicated by the blue regions). Evidently, multiple combinations of localization and inflation values yield comparable analysis errors. Furthermore, the tuning plots appear nearly identical across the three observation networks, suggesting that LETKF’s tuning is largely insensitive to changes in the spatial distribution of observations. This is further supported by the fact that the optimal localization and inflation values remain similar across all three networks.

The second column of Figure \ref{fig3_letkf_tuning} demonstrates that introducing even a small fraction (5\%) of nonlinear observations significantly reduces the range of localization and inflation parameters for which LETKF remains stable and avoids filter divergence. As an extreme case, in the RANDOM network (Figures \ref{fig3_letkf_tuning}g-i) with 5\% nonlinearity, only a single parameter combination (a localization scale of $1000$ km and RTPS of $0.9$) results in a well-behaved LETKF algorithm with sufficiently small RMSEs. As the fraction of nonlinear observations increases to 15\%, tuning becomes even more difficult. For the RANDOM network, no combination of localization and inflation values prevents LETKF from experiencing large analysis RMSEs and filter divergence cases.

A completely different behavior emerges in the EnSF experiments. As shown by the solid curves in Figure \ref{fig2_rmse_cycle}, the RMSE time series exhibit minimal variability, both with respect to the fraction of nonlinear observations and the overall spatial distribution of the assimilated network. This robustness persists despite using a single version of EnSF that is neither localized nor tuned for optimal inflation values \citep[note the EnSF configuration is the same as in][]{EnSF_SQG}.

\begin{figure}[ht!] 
\noindent \includegraphics[width=\textwidth]{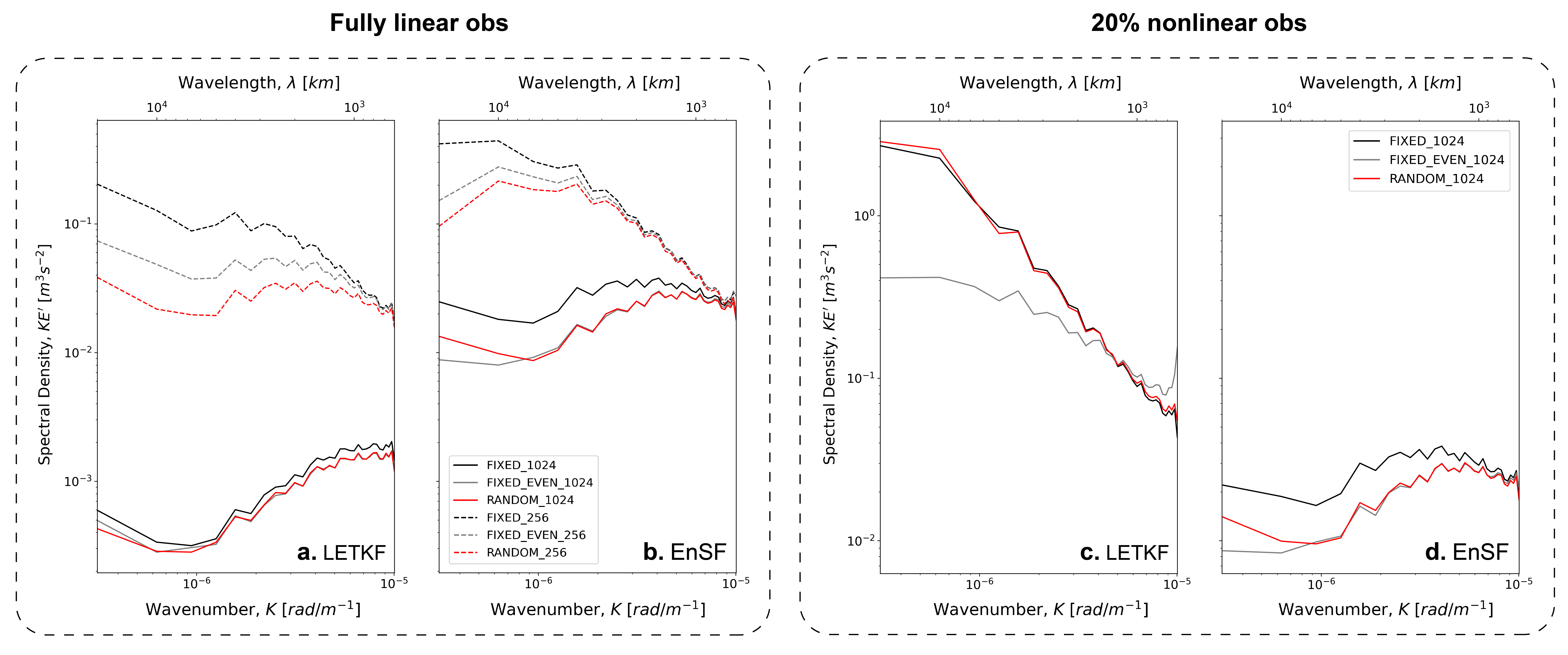}
\caption{Comparison of the multiscale impacts on LETKF and EnSF analyses through the kinetic energy spectrum of the analysis mean errors for a fully linear observation network (left dashed box) and a network containing 20\% nonlinear (arctangent) observations (right dashed box). Solid (dashed) lines correspond to experiments assimilating 1024 (256) observations, respectively. The color of each spectral density curve indicates the spatial distribution of the assimilated network.}
\label{fig4_ke_spectra}
\end{figure}

Since the SQG model encompasses multiple scales of motion, it is worth analyzing how different observation networks influence the spectral characteristics of the analysis mean errors. The left panel of Figure \ref{fig4_ke_spectra} focuses on the scenario in which the entire observation network consists of linear observations.

Examining the LETKF results first, we confirm the expected trend that increasing the number of observations from 256 ($6.25\%$ of all state variables) to 1024 ($25\%$ of all state variables) induces lower analysis errors across all scales. However, the error reduction is most pronounced at the smallest wavenumbers (largest wavelengths), resulting in a modification of the overall spectral slope. In particular, the sparser observation network leads to more substantial analysis errors at large scales. According to the numerical experiments of \citet{durran_gingrich_2014}, this scenario is more conducive to a rapid growth of the subsequent forecast errors. The effect arises from the forward cascade of energy in a turbulent regime characterized by a -5/3 power law, which applies to the SQG dynamics simulated in this study. Conversely, the denser networks with 1024 observations correspond to the opposite case: most of the analysis errors are concentrated within the small scales, which would imply a slower forecast error saturation. 

Similar effects are observed in the EnSF experiments, although the overall sensitivity to the number of observations is smaller than LETKF. For example, the spectral density happens to be nearly identical at the smallest wavelengths across all networks.

The impacts caused by the spatial distribution of observations also reveal several insightful findings. For example, differences among FIXED, FIXED\_EVEN, and RANDOM in LETKF (Figure \ref{fig4_ke_spectra}) become more pronounced in the sparse observation regime. By contrast, the EnSF analysis errors exhibit clearer separation in the denser network case, particularly at larger scales. Nevertheless, a consistent finding across both filters is that the FIXED network yields the largest analysis errors regardless of how many observations are assimilated.

The spectral decomposition of the LETKF and EnSF analysis errors with 20\% nonlinear (arctangent) observations is shown in panels (c) and (d) of Figure~\ref{fig4_ke_spectra}. Results are presented only for networks with 1024 observations (25\% of all state variables) due to the divergent behavior of LETKF in sparser regimes. Consistent with previous findings, we observe minimal differences in the EnSF analysis results when nonlinear observations are assimilated. On the other hand, the nonlinear LETKF experiments exhibit stark differences in the magnitude of analysis errors as a function of scale. The resulting spectral density curves resemble those in the sparser observation regime shown in Figure~\ref{fig4_ke_spectra}a, with the most substantial analysis errors occurring at the largest scales of motion. Thus, in addition to the overall degradation of the LETKF performance, the slope of the kinetic energy spectrum would favor a more rapid error growth in subsequent SQG predictions \citep{durran_gingrich_2014}.

\section{Conclusions}
\label{sec:conclusions}

This study explored the sensitivity of different ensemble filters to variations in the assimilated observation networks. Numerical experiments were conducted using the surface quasi-geostrophic (SQG) system, which exhibits turbulent characteristics similar to those of real geophysical flows. We examined two DA methods: the standard Local Ensemble Transform Kalman Filter (LETKF) and the recently developed Ensemble Score Filter (EnSF), which leverages diffusion models to sample from complex analysis distributions.

We compared both ensemble filters with respect to several key aspects of the assimilated observation networks, including density, spatial distribution, and the fraction of nonlinear observations in the system. A common finding across the two DA methods was that a fixed network with evenly distributed observations (FIXED\_EVEN) resulted in smaller analysis errors compared to an inhomogeneous network (FIXED), which is more representative of real Earth observing systems. By contrast, a network of randomly distributed observations (RANDOM) exhibited different impacts depending on the chosen DA method. For the non-Gaussian EnSF, it yielded analysis skill comparable to the well-performing FIXED\_EVEN configuration, while the Gaussian LETKF often performed worse relative to the other two networks.

Although it is well known that increasing the number of observations improves the analysis quality, our experiments provided new insight into how observation density alters the spectral characteristics of the analysis errors. Notably, we observed changes in the slope of the kinetic energy spectrum, which has implications for the growth of initial condition uncertainty \citep{durran_gingrich_2014,rotunno_snyder_2008}.

Gradually increasing the fraction of nonlinear observations also led to important consequences. While an optimally tuned LETKF performed well for a fully linear observation network, the benefits of the more flexible EnSF became apparent even when only a small fraction of the assimilated observations were nonlinear (5\% in the RANDOM network case). The presence of nonlinear observations was also shown to alter the spectral characteristics of the LETKF analysis errors, leading to visible increases at the large scales. On the contrary, EnSF maintained consistent performance across all network configurations, despite the absence of explicit localization or inflation tuning.

There are several practical implications of our results for operational NWP systems, as many of them rely on variants of the EnKF. For example, the German Kilometer-scale ENsemble-based Data Assimilation (KENDA) system \citep{schraff_et_al_2016} is based on the LETKF approach to provide convective-scale DA analyses. Our LETKF tuning experiments (Figure~\ref{fig3_letkf_tuning}) suggest that it becomes increasingly more difficult to identify optimal localization and inflation parameters when nonlinear observations are assimilated. According to personal correspondence with KENDA developers, radar reflectivity data account for just over 20\% of the observations in their DA system, based on statistics from February 11, 2025. These observations are highly nonlinear and conceptually resemble the arctangent observation operator used in our study. Although their current DA system remains stable, the projected growth in other nonlinear observations (e.g., all-sky radiances) may challenge LETKF's effectiveness in assimilating such data. This highlights the need for additional case studies evaluating the performance of LETKF and other EnKF variants in highly nonlinear observational regimes.

The reduced sensitivity of EnSF to both the spatial distribution of observation networks and the degree of nonlinearity (e.g., Figure~\ref{fig2_rmse_cycle}) raises the broader question of whether such robustness is a general property of other nonlinear DA methods. A systematic exploration of this topic in realistic geophysical systems like KENDA would be of practical value, as it will offer guidance for the design of future observation networks. Moreover, the development of new impact metrics that more accurately account for nonlinear effects will be necessary to better evaluate and prioritize these emerging observation types.

\normalsize
\subsubsection{Open-source code access} Python scripts for running the ensemble filtering experiments with the SQG model and for reproducing the figures are available on GitHub (\url{https://github.com/ZixiangXiong/sqgpublic}) and Zenodo (\url{https://zenodo.org/records/15313555?token=eyJhbGciOiJIUzUxMiJ9}).

\normalsize
\subsubsection{Acknowledgments} Feng Bao would like to acknowledge the support from U.S. National Science Foundation through project DMS-2142672 and the support from the U.S. Department of Energy, Office of Science, Office of Advanced Scientific Computing Research, Applied Mathematics program under Grants DE-SC0025412. The corresponding author, Hristo Chipilski, would like to acknowledge support from Florida State University's CRC Seed Grant 047080. He also expresses gratitude to Jeffrey Whitaker for providing the SQG model and LETKF codes, and to him, along with Laura Slivinski, Hendrik Reich, and Harald Anlauf, for subsequent discussions regarding the practical significance of the EnSF results for real observation networks. In addition, Hristo Chipilski appreciates the valuable feedback by Xuguang Wang, David Parsons, Alan Shapiro, and Evgeni Fedorovich on an earlier version of the reference LETKF experiments presented in this study.

\newpage
\renewcommand\bibname{References}
\bibliographystyle{ametsoc2014}
\bibliography{manuscript}

\end{document}